\documentclass[aps,prb,preprint,amsmath,amssymb,showpacs]{revtex4}
\usepackage{graphicx}
\usepackage{dcolumn}
\usepackage{bm}

\begin{document}

\title{Realization, Characterization, and Detection of Novel Superfluid Phases with Pairing between Unbalanced Fermion Species}

\author{Kun Yang}

\affiliation{National High Magnetic Field Laboratory and Department
of Physics, Florida State University, Tallahassee, Florida 32306,
USA}

\date{\today}

\begin{abstract}
In this chapter we review recent experimental and theoretical work
on various novel superfluid phases in fermion systems, that result
from pairing fermions of different species with unequal densities.
After briefly reviewing existing experimental work in
superconductors subject to a strong magnetic field and trapped cold
fermionic atom systems, we discuss how to characterize the possible
pairing phases based on their symmetry properties, and the
structure/topology of the Fermi surface(s) formed by the unpaired
fermions due to the density imbalance. We also discuss possible
experimental probes that can be used to directly detect the
structure of the superfluid order parameter in superconductors and
trapped cold atom systems, which may establish the presence of some
of these phases unambiguously.

\end{abstract}

\maketitle

\section{Introduction and brief review of experimental work}

It is well known that superfluidity and superconductivity in
fermionic systems result from pairing of fermions, and the Bose
condensations of these so-called Cooper pairs.  In a specific
fermion system, Cooper pairs are often made of fermions of different
species; for example in superconductors they are electrons of
opposite spins. Thus the most favorable situation for pairing is
when the two species of fermions have the same density, so that
there is no unpaired fermion in the ground state. The physics of
pairing and resultant superfluidity under such condition is well
described by the highly successful Bardeen-Cooper-Schrieffer (BCS)
theory. It has been a long-standing fundamental question as to what
kind of pairing states fermions can form when the two fermion
species have different densities. A closely related issue is that in
any paired or superfluid state formed under such situation, some of
the majority fermions will necessarily be unpaired; thus a related
question is how the system accommodates these unpaired fermions. An
early suggestion was due to Fulde and Ferrell\cite{ff}, and Larkin
and Ovchinnikov\cite{lo}, who argued that the Cooper pairs may
condense into either a single finite momentum state, or a state that
is a superposition of finite-momentum states. Such states are known
collectively as the Fulde-Ferrell-Larkin-Ovchinnikov (FFLO) state;
they break translation and rotation symmetries. More recently other
suggestions have been put forward, including deformed Fermi surface
pairing (DFSP)\cite{ms,ms1} and breached pairing
(BP)\cite{liu,glw,forbes} states, each with their distinct symmetry
properties.

Experimentally, the issue of unbalanced pairing arise in several
different contexts. Historically, it first arose in the context of a
singlet superconductor subject to a large Zeeman splitting. The
Zeeman splitting could be due to either a strong external magnetic
field, or an internal exchange field (in the case of a ferromagnetic
metal/superconductor). Under such a strong magnetic or exchange
field, there is a splitting between the Fermi surfaces of spin-up
and -down electrons. The original FFLO proposal was advanced in this
context. However the FFLO state has not been observed in
conventional low-$T_c$ superconductors. The reason for that, we
believe, is because these superconductors are mostly
three-dimensional, and the magnetic field that gives rise to Zeeman
splitting also has a very strong orbital effect, which suppresses
the superconductivity before the Zeeman effect becomes significant.
The situation has changed recently, as experimental results
suggestive of the FFLO state in heavy-fermion, organic, and
high-$T_c$ superconductors have been found\cite{gloos, tachiki,
modler1996a, modler1996b, brooks, singleton, balicas,tanatar}. These
compounds are quasi-one or quasi-two-dimensional, thus the orbital
effect is weak when the magnetic field is aligned in the conducting
plane or along the chain; as a consequence the upper critical field
of the superconductor is comparable to or exceeds the so-called
Pauli paramagnetic limit\cite{clogston}, a field at which the Zeeman
splitting becomes comparable to the superconducting gap. The the
FFLO state becomes possible in such cases. Recent experimental
results in CeCoIn$_5$, a quasi-2D $d$-wave superconductor, are
particularly encouraging~\cite{radovan, bianchi2003, capan,
watanabe, martin, kakuyanagi, bianchi2005}, as various experimental
probes, including specific heat\cite{radovan, bianchi2003}, thermal
conductivity\cite{capan}, ultrasound\cite{watanabe}, penetration
depth\cite{martin} and NMR\cite{kakuyanagi} all identify a novel new
phase in a region of the temperature-magnetic field phase diagram
where the FFLO phase is expected theoretically. This system is by
far the most promising candidate for the realization of the FFLO
phase in superconductors at this point.

More recently, fermion pairing and superfluidity have become the
focus of experimental work on trapped cold atom systems. Compared to
electronic superconductors, one big advantage of such systems is
that the strength of the pairing interaction can be very well
controlled by manipulating the so-called Feshbach resonance, and one
can explore a wide range of interaction strength from the weak
coupling BCS regime to the strong coupling regime in which pairs of
fermionic atoms form closely bound bosonic molecules (the so-called
BEC regime). Another, perhaps more important advantage, is that
experimentalists can induce and control the imbalance by simply
mixing atoms of different species with different numbers. In
contrast the imbalance in superconductors (Zeeman splitting) is due
to an external magnetic field; the field, however, also brings the
orbital effect that complicates the situation considerably. Indeed,
in the very first such experiments unequal numbers of two hyperfine
states of fermionic $^6$Li atoms were mixed and scanned across a
Feshbach resonance.\cite{zwierlein,randy} In these experiments it
was found that paired and unpaired fermions phase
separate\cite{bedaque} when the imbalance is large. In one of the
experiments\cite{randy} it was found that the fermions do {\em not}
phase separate when the imbalance is sufficiently small; further
experiments are needed to clarify the nature of the state in such a
situation.

Another place where pairing between unbalanced fermion species
arises is quark and nucleon pairing in high density quark or nuclear
matter, such as in the core of a neutron star. There the origin of
density imbalance is due to the difference in the rest mass of
quarks or nucleons that form the pairs; when the different pairing
species are in chemical equilibrium (meaning they have the same
chemical potential), their Fermi momenta and therefore densities are
different. The physics of quark and nucleon pairing have been
previously reviewed in Ref. \onlinecite{casalbuoni}, and is also
covered with great detail in other chapters of this volume.

\section{Characterization of Phases based on Symmetry and topology
of Fermi surface(s)}

As discussed in Section I, a number of possible phases have been
proposed theoretically in fermionic superfluids with unbalanced
pairing species. The purpose of this section is to classify these
phases based on their symmetry and other properties, and discuss the
relations between these phases based on such classification. This
also gives us insight into the nature of the phase transitions
between various phases. We note that classification (or
characterization) of classical phases and phase transitions are
based on Landau theories, whose forms are completely determined by
the symmetry properties of the phases involved; thus our
classification based on symmetry considerations are complete at
finite temperature ($T$). It has been realized recently, however,
that such classification may not be complete for quantum phases and
phase transitions\cite{wen}; additional classification schemes may
be necessary at $T=0$. Here we propose that in these pairing phases
the structure and in particular, topology of the Fermi surfaces
formed by unpaired fermions can be used as additional classification
scheme to characterize phases and phase transitions. Most of the
ideas behind such considerations were originally presented in Refs.
\onlinecite{yang06,ys,sy}, which we review below.

We begin with symmetry considerations, and start our discussion from
the FFLO state, which has the longest history of studies. Following
the superconductivity terminology, throughout the rest of this
chapter we will use ``spin" indices $\sigma=\uparrow, \downarrow$ to
label the two different species of fermions that form Cooper pairs,
``Zeeman splitting" $\Delta\mu=\mu_\uparrow-\mu_\downarrow$ to
represent the chemical potential difference between the two fermion
species that form Cooper pairs, and ``magnetization" $m$ to
represent their density difference. When $\Delta\mu\ne 0$, up- and
down-spin electrons form Fermi seas with different Fermi momenta
$p_{F\uparrow}$ and $p_{F\downarrow}$ in the normal state; it was
thus suggested\cite{ff,lo} that when pairing interaction is turned
on, the initial pairing instability is for fermions with opposite
spins on their respective Fermi surfaces to pair up and form a
Cooper pair with a net momentum $p\approx
p_{F\uparrow}-p_{F\downarrow}$. This results in a pairing order
parameter $\Delta({\bf r})$ that is oscillatory in real space, with
period $2\pi\hbar/p$. In general the structure of $\Delta({\bf r})$
is characterized not just by a single momentum $p$, but also by its
higher harmonic components. More detailed mean-field
study\cite{rainer} suggested the following real space picture for
FFLO state: it is a state with a finite density of uniformly spaced
domain walls; across each domain wall the order parameter $\Delta$
(which is real in the mean-field theory) changes sign, and the
excess magnetization due to spin imbalance are localized along the
domain walls, where $\Delta$ (which is also the gap for unpaired
fermions) vanishes; see Fig. 1a of Ref. \onlinecite{yang06} for an
illustration\cite{note}. Thus the total magnetization is
proportional to the domain wall density. This picture was made more
precise by an exact solution in one-dimension (1D) based on
bosonized description of spin-gapped Luttinger liquids\cite{yang01}
(also refered to as the Luther-Emery liquid in condensed matter
literature), where the domain walls are solitons of the sine-Gorden
model that describes the spin sector; each soliton carries one
half-spin. While quantum and thermal fluctuations do not allow true
long-range order in 1D, such order can be stabilized by weak
interchain couplings\cite{yang01}. Coming back to isotropic high D
cases, it is clear that the presence and ordering of these domain
walls break rotation symmetry, and translation symmetry in the
direction perpendicular to the walls, although translation symmetry
along the wall remains intact. Thus the symmetry properties of the
FFLO state is identical to that of the smectic phase of liquid
crystals (smectic-A phase to be more precise)\cite{degennes}.

Once the mean-field FFLO state is identified with the smectic phase
of liquid crystals based on symmetry considerations, one can borrow
insights as well as known results on the thermodynamic phases of
liquid crystals to the present problem\cite{yang06}. In classical
liquid crystals it is known that as one increases thermal
fluctuations, the broken symmetries of the smectic phase are
restored in the following sequence\cite{degennes}: the translation
symmetry is restored first when the smectic melts into a nematic
that breaks the rotation symmetry only, and then the nematic melts
into an isotropic liquid that has no broken spatial symmetry. We
thus expect the same sequence of phases and phase transitions occur
in superfluids with unbalanced fermion pairing, as we increase the
strength of either thermal or quantum fluctuations. Very
interestingly, the nematic and isotropic phases have precisely the
same symmetry properties as the deformed Fermi sea pairing
(DFSP)\cite{ms,ms1} and breached pairing (BP)\cite{liu,glw,forbes}
states, which were proposed as alternative phases for unbalanced
pairing that compete with the FFLO phase. In the DFSP phase, the
(originally mismatched) Fermi surfaces of the majority and minority
fermion species deform spontaneously, so that they match in certain
regions in momentum space to facilitate pairing; the rotation
symmetry is broken by the Fermi surface distortion, but the
translation symmetry remains intact (see Fig. 2 of Ref.
\onlinecite{ms} for an illustration). In the BP phase, an isotropic
shell in momentum space is used to accommodate the excess
magnetization, while pairing occurs in the rest of the momentum
space; both rotation and translation symmetries are intact in this
phase. It should be noted that the variational states studied in
Refs. \onlinecite{ms,ms1,liu,glw,forbes} are quite simple and
essentially of mean-field type; they look quite different from the
real space picture developed in Ref. \onlinecite{yang06} (see its
Fig. 1) based on considerations of fluctuation effects. We would
like to emphasize, however, that it is the common symmetry
properties that allowed us to identify the DFSP state as the nematic
phase, and the BP state as the istropic phase of the liquid crystal.
The symmetry considerations also suggests a unified understanding of
all three of these states as different phases of a liquid crystal.
We note in passing that very similar considerations have also led to
deeper understanding of different phases in cuprate
superconductors\cite{kfe} and quantum Hall
liquids\cite{fk,radzihovsky}.

Coming back to the smectic state, one should in principle also
consider the possibility that further symmetry breaking occurs along
directions parallel to the domain walls; this will result in
breaking of translation symmetries in {\em all} directions, and
result in a crystal version of the FFLO state\cite{note}.

As usual the strength of thermal fluctuation is controlled by
temperature ($T$). The quantum fluctuations (QF), on the other hand,
are controlled by the strength of pairing interactions; QF is weaker
for weak pairing interactions (the BCS regime, where the superfluid
state is well described by the mean-field theory), while stronger
for strong pairing interaction (the BEC regime). In Ref.
\onlinecite{yang06} a phase diagram has been proposed based on such
considerations, in which an infinitesimal density imbalance is
present (assuming there is no phase separation in this case, and the
smectic phase is the least symmetric phase that is realized), while
both temperature and pairing interaction strength are varied. As
discussed earlier, the pairing interaction strength can be
controlled by manipulating the Feshbach resonance in trapped cold
atom systems; thus one may be able to explore the entire phase
diagram in such systems.

At finite temperature, all phases and phase transitions are
classical, and are fully characterized by symmetries. We thus
conclude that our classification of the possible pairing phases
based on symmetry is complete, and the crystal (FFLO), smectic
(FFLO), nematic (DFSP) and isotropic (BP) phases discussed above
exhaust all possible phases in this case. Furthermore, symmetry also
dictates the nature of the phase transitions. Again, based on known
results from studies of classical liquid crystals\cite{degennes}, we
expect the transition between nematic (DFSP) and isotropic (BP)
phases to be generically first order, while the transition between
nematic (DFSP) and smectic (FFLO) phases is most likely 2nd order. A
direct transition between  FFLO (either crystal or smectic) and
isotropic (BP) phases is unlikely; should such a transition occur,
it will be first order.

At zero temperature, the possible phases are characterized by the
ground state of the system, and the low-lying excitations above it,
which are intrinsically quantum mechanical. It has become
increasingly clear in recent years that characterization of quantum
phases based on symmetry alone is often insufficient, and additional
characterization schemes are needed to classify such ``quantum" or
``topological" order\cite{wen}. At present we do not yet have a
complete classification scheme for quantum order\cite{wen}. In the
following we will argue that in the problem of unbalanced pairing
discussed here, one may use the properties of the Fermi surface(s)
formed by unpaired fermions, especially their topology, to
characterize all the possible phases; combined with symmetry
properties discussed above, they most likely provide a complete
classification scheme. We note that Fermi surfaces are sharp and
well-defined objects only at $T=0$; finite $T$ smears Fermi
surfaces, and as a consequence they are no longer well-defined.

It is intuitively clear that in the presence of imbalance, some of
the majority fermions will be unpaired; these unpaired fermions will
form a Fermi sea of its own with at least one, but possibly more
Fermi surfaces. Recently this intuitive picture has been made
quantitatively precise in the form of a mathematically rigorous
theorem\cite{sy}, which is a generalization of the Luttinger's
theorem\cite{luttinger} for normal metals to the case with pairing
interaction and superfluidity. The theorem makes distinction between
two different cases\cite{sy}:

(i) In the absence of superfluidity, there are two Fermi surfaces
for spin-up and -down fermions, whose volumes are {\em individually}
conserved:
\begin{equation}
N_\uparrow = \frac{A}{(2 \pi)^d} \Omega_\uparrow~~, ~~ N_\downarrow
= \frac{A}{(2 \pi)^d} \Omega_\downarrow , \label{lutt2fs}
\end{equation}
where $d$ is the dimensionality, $A$ is the (real space) volume of
the system, and $\Omega$ is the (momentum space) volume enclosed by
the Fermi surface. We emphasize that this is an exact result that
applies even when the pairing interaction is so strong that some of
the fermions may form very closely bound pairs or ``molecules"; in
this case one might intuitively expect that these fermions in
closely bound states would not contribute to the Fermi surface
volume. Our result indicate that as long as there is no
superfluidity (or the pairs do not condense), it is the {\em total}
numbers of fermions that dictate the volumes of Fermi surfaces.

(ii) In the presence of superfluidity, or when Cooper pairs Bose
condense and the U(1) symmetry associated with charge conservation
is spontaneously broken, the spin-up and -down Fermi surface volumes
are no longer individually conserved. However their difference
remains to be conserved, and is dictated by the imbalance:
\begin{equation}
\Delta N=N_\uparrow-N_\downarrow={A\over
(2\pi)^{d}}(\Omega_\uparrow-\Omega_{\downarrow}). \label{lutt1}
\end{equation}
In this case we can have either one or two Fermi surfaces; when
there is only one Fermi surface we simply have
$\Omega_{\downarrow}=0$.

In our discussion so far we have assumed the system to be uniform or
translationally invariant. These results, however, can be
generalized to cases with spontaneously broken translational
symmetry, which is the case for the FFLO state. In such cases, the
Fermi surface volumes are well-defined modulo the Brillouin zone
volume $\Omega_B$; as a consequence all of our statements on the
constraints on Fermi surface volumes are modulo $\Omega_B$. The
situation is identical to electrons moving in a periodic potential
considered by Luttinger originally \cite{luttinger}.

The theorem discussed above, in particular the constraint of Eq.
(\ref{lutt1}), dictates that the ground states of the systems
considered here can be characterized by their Fermi surfaces, and
there must be gapless quasiparticle excitations near these Fermi
surfaces. We can thus use the Fermi surfaces as an additional
classification scheme for the unbalanced pairing phases at $T=0$,
and expect the following generic cases:

(i) One Fermi surface for spin-up fermions. In this case its volume
is fixed to be
\begin{equation}
\Delta N=N_\uparrow-N_\downarrow={A\over (2\pi)^{d}}\Omega_\uparrow.
\end{equation}

(ii) Two Fermi surfaces, whose volumes are not fixed individually,
but their difference are fixed by Eq. (\ref{lutt1}).

(iii) No Fermi surface. In this case  Eq. (\ref{lutt1}) indicates
$\Delta N=0$, so there is no imbalance.

We believe combining the symmetry property (crystal, smectic,
nematic or isotropic) with the number of Fermi surfaces, we have an
essentially complete characterization of all the possible quantum
pairing phases. As an example, we expect two possible isotropic
phases. The breached pair (BP) phase (also known as Sarma
phase\cite{sarma}), in its original form\cite{liu}, has two Fermi
surfaces, which may be stable at weak coupling. On the other hand we
expect an isotropic phase with a single Fermi surface at strong
coupling\cite{yang06,son,sheehy}. For the nematic case, we can in
principle again have one or two Fermi surfaces; in this case the
pairing order parameter is uniform and does not break any spatial
symmetry, while the Fermi surface(s) should be anisotropic and break
rotation symmetry spontaneously. As already mentioned, in the FFLO
phase, due to the broken translation symmetry, Brillouin zones form
and the Fermi surface(s) are folded into a Brillouin zone.

The transitions between different phases with different numbers of
Fermi surfaces have been discussed in Refs. \onlinecite{ys,sy}.

\section{Detection of Phases based on ``Phase Sensitive" Experimental Probes}

As discussed in the previous section, the possible phases for
systems with pairing between unbalanced fermion species can be
characterized by (i) their symmetry properties, especially those
associated with the spatial structure of the pairing order
parameter; and (ii) in the case $T = 0$, the structure and in
particular, topology of the Fermi surfaces formed by unpaired
fermions. Thus to experimentally identify a phase unambiguously, one
needs to have experimental methods that probe (i) and/or (ii)
directly. While there have been quite a few experiments that study
possible FFLO phases in various systems, and the studies of
CeCoIn$_5$ are getting more and more detailed, none of the existing
experiments probes (i) or (ii) directly. In the following we will
discuss a few possible experiments that probe either (i) or (ii), in
either electronic superconductors or trapped cold atom systems.

\subsection{Detecting Spatial Structure of Pairing Order Parameter
in Superconductors Using Phase Sensitive Experimental Probes}

In this subsection we will briefly discuss three possible
experimental methods that directly probe the spatial structure of
the pairing order parameter, in the FFLO state of electronic
superconductors, which we have considered
recently\cite{kun2000,ym,cui}. We also discuss a proposed
experiment\cite{bbm} that probes physics similar to that of Ref.
\onlinecite{kun2000}, as well as the possibility of using neutron or
muon scattering to detect the spin structure of the FFLO state.

{\bf Josephson Effect between a BCS and an FFLO superconductor} ---
In Ref. \onlinecite{kun2000} we demonstrated that one can use the
Josephson effect between an FFLO superconductor and a BCS
superconductor to measure the momenta of (in principle) all the
Fourier components of the pairing order parameter of the FFLO
superconductor. The idea behind this proposal is quite simple.
Consider a two-dimensional BCS superconductor, described by a
spatially dependent superconducting order parameter $\Psi_{BCS}({\bf
r})$, which is coupled to a two-dimensional FFLO superconductor,
described by an order parameter $\Psi_{FFLO}({\bf r})$. We consider
the two Josephson junction geometries shown in Figure 1 of Ref.
\onlinecite{kun2000}. Since the physics for the two geometries are
similar we focus our discussion on geometry of Fig. 1a of Ref.
\onlinecite{kun2000}, in which the two superconductors are stacked
on top of each other. In the Ginsburg-Landau description, the
Josephson coupling term in the free energy takes the form (in the
absence of any magnetic flux going through the junction, or in
between the two superconductors)
\begin{equation}
H_J=-t\int{d^2{\bf r}}[\Psi_{FFLO}^*({\bf r})\Psi_{BCS}({\bf r})+
c.c.], \label{eq1}
\end{equation}
where t is the Josephson coupling strength. In the ground state of a
BCS superconductor, $\Psi_{BCS}({\bf r})= \psi_0$ is a constant.
However, in an FFLO superconductor the order parameter is a
superposition of components carrying finite momenta:
\begin{equation}
\Psi_{FFLO}({\bf r})=\sum_{m}\psi_m e^{i{\bf k}_m\cdot{\bf r}},
\end{equation}
and is oscillatory in space. In the absence of magnetic flux inside
the junction, the total Josephson current is
\begin{equation}
I_J = {\rm Im}\left[t \int{d^2{\bf r}}\Psi^*_{BCS}({\bf
r})\Psi_{FFLO}({\bf r})\right] = \sum_m {\rm Im}
\left[t\psi_0^*\psi_m\int{d^2{\bf r}}e^{i{\bf k}_m\cdot{\bf
r}}\right].
\end{equation}
Clearly, due to the oscillatory nature of the integrand, the
Josephson current is suppressed in such a junction.

Mathematically, the reason that the Josephson current is suppressed
here is similar to the suppression of Josephson current by an
applied magnetic field in an ordinary Josephson junction between two
BCS superconductors. However, the physics is very different: here
the suppression is due to the spatial oscillation of the {\em order
parameter} in the FFLO state, while in the case of ordinary
Josephson junction in a magnetic field, the phase of the Josephson
tunneling {\em matrix element} is oscillatory (in a proper gauge
choice). Nevertheless, the mathematical similarity allows these two
effects to {\em cancel} each other and restore the Josephson
current, by applying an appropriate amount of magnetic flux through
the junction, and the amount of flux that restores the Josephson
effect is a direct measure of the momentum of one of the Fourier
components of the pairing order parameter of the FFLO
superconductor. This was demonstrated in Ref. \onlinecite{kun2000},
and we refer the reader to this paper for detailed analyzes using
both the effective Ginsburg-Landau description and microscopic
theory, as well as for an alternative geometry. This idea has some
similarity to the so called ``phase sensitive" experiments that
unambiguously determined the d-wave nature of the pairing order
parameter of high T$_c$ cuprate
superconductors\cite{harlingen,tsuei}. However unlike the cuprate
experiments that attempt to determine the {\em internal} structure
of the Cooper pairs (or their anugular momentum), here\cite{kun2000}
we use the sensitivity of the Josephson coupling to the phase of the
pairing order parameter to determine its {\em spatial} structure, or
the momentum of the Cooper pairs.

In the following we discuss a few practical issues that may arise
when trying to implement this proposal experimentally.

(i) We want to use the Josephson effect to probe the spatial
structure of the pairing order parameter of the FFLO superconductor,
using the BCS superconductor (whose pairing order parameter is
uniform in space) as a reference point. In order for this idea to
work however, the BCS and FFLO superconductors should have the same
{\em internal} structure for their pairing order parameter, i.e.,
the two superconductors should be both s-wave or both d-wave etc,
otherwise the Josephson current will vanish simply due to the
mismatch in internal symmetry. As noted earlier, the most promising
candidate for FFLO state thus far is CeCoIn$_5$, which is a d-wave
superconductor. Thus to implement this idea on CeCoIn$_5$ one needs
to use another d-wave superconductor for the reference BCS state. A
natural choice is thus a cuprate superconductor, which has the
additional advantage that it has a much bigger gap and higher Pauli
limit than CeCoIn$_5$; thus when placed in a strong magnetic field
(about 10T, necessary to drive CeCoIn$_5$ into the FFLO state), it
is still in the BCS phase.

(ii) The key ingredient that makes the Josephson effect useful in
the determination of the structure of the FFLO pairing order
parameter is that one needs to adjust the magnetic flux in the
junction to have the Josephson effect; the order parameter momentum
can be determined from the magnetic flux. On the other hand we also
need to put the superconductors in a strong magnetic field to drive
one of them into the FFLO phase, unless it is a ferromagnetic
superconductor that has a spontaneous magnetization. Thus the
magnetic field that stabilizes the FFLO state may interfere with the
flux through the junction. The configuration that avoids this
complication is the one depicted in Fig. 1b of Ref.
\onlinecite{kun2000}, in which the BCS and FFLO superconductors,
both assumed to be (quasi) two-dimensional, are placed side-by-side.
The advantage of this configuration is that the magnetic field that
stabilizes the FFLO state is an {\em in-plane} magnetic field, which
does not contribute to the flux through the junction that controls
the Josephson effect. As a result the in-plane field and (out of
plane) flux through the junction can be tuned independently.

(iii) In an infinite system, which was analyzed in Ref.
\onlinecite{kun2000}, the Josephson current is exactly zero unless
the relative phase oscillation between the BCS and FFLO
superconductors are canceled exactly by the phase oscillation in
Josephson coupling induced by the flux through the junction. In real
systems the junction has a finite size; we thus expect a Fraunhofer
pattern in the flux-dependence of the Josephson current, which is
peaked at a finite flux strength determined by the momentum of
pairing order parameter of the FFLO superconductor.

{\bf Exotic Vortex Structure of FFLO Superconductors} --- The FFLO
state is stabilized by the Zeeman effect of an external magnetic
field. On the other hand the field can also generate an orbital
effect; for example in a purely 2D system, the Zeeman effect is
determined by the total magnetic field, while the orbital effect is
generated by the out-of-plane component of the magnetic field, when
it is nonzero. Thus the relative importance between the Zeeman and
orbital effect can be controlled by the angle between the magnetic
field and the 2D plane. The orbital effect generates vortex states,
which can be used to detect FFLO physics. The idea here goes back to
an early observation by Bulaevskii, who pointed out that \cite{bul}
depending on the interplay between the orbital and Zeeman effects of
the magnetic field, the order parameter of a FFLO state near its
upper critical field can correspond to a high Landau level (LL)
index Cooper pair wave function. Recent
work\cite{sr,krs,hb,hbbm,klein} on the FFLO vortex lattice structure
(VLS) in specific situations has demonstrated that these high LL
index VLS's can be very different from the triangular lattice
Abrikosov VLS favored by lowest Landau level (LL) Cooper pairs.
However it remained a challenging task to determine the vortex
lattice structure for FFLO superconductors under general conditions.
The difficulty has its origin in the complicated Ginsburg-Landau
theory appropriate for FFLO superconductors\cite{kulic}:
\begin{equation}
F\propto|(-\nabla^2-q^2)\psi|^2+a|\psi|^2+b|\psi|^4+c|\psi|^2|\nabla\psi|^2
+d[(\psi^*)^2(\nabla\psi)^2+\psi^2(\nabla\psi^*)^2]+e|\psi|^6+\cdots,
\label{F}
\end{equation}
where $a,b,c,d,e$ and $q$ are parameters that depend on both
temperature and Zeeman splitting. The fundamental difference between
FFLO and BCS superconductors is expressed by the first term in $F$
which describes the {\em kinetic energy} of the order parameter; in
an FFLO superconductor this term is minimized when the order
parameter carries a finite wave vector (or momentum) $q$. Thus far
we have only taken into account the Zeeman effect of the external
magnetic field; for a 2D superconductor with the field
 ${\bf B}$
tilted out of system plane, orbital coupling must be accounted for
by performing a minimal substitution $\nabla\psi\rightarrow {\bf
D}\psi = (\nabla-2ie{\bf A}/c)\psi$ with $\nabla\times {\bf
A}=B_{\perp} \hat{z}$. This leads to Landau quantization of the
kinetic energy term, namely the eigenvalues of ${\bf
D}^2=(\nabla-2ie{\bf A}/c)^2$ are $-(2n+1)/\ell^2$, where
$\ell=\sqrt{\hbar c/2eB}$ is the Cooper pair magnetic length and
$n=0, 1, 2,\cdots$ is the LL index.

There are two specific sources of difficulty, compared to the vortex
states of a BCS superconductor, that results in the Abrikosov
lattice. (i) For a BCS superconductor the kinetic energy is
minimized by $n=0$, {\em i.e.}, $\psi$ is a lowest LL wave function.
For an FFLO superconductor, however, the kinetic energy is minimized
by the index $n$ that minimizes $|(2n+1)/\ell^2-q^2|$. This can lead
to high Landau level wave functions which are much more complicated.
(ii) For the BCS case, one only needs to minimize the $|\psi|^4$
term in Eq. (\ref{F}). For FFLO superconductors however, due to the
fact that the order parameter carries a finite momentum, there are
additional quartic terms (which involve spatial gradients) that make
substantial contribution to the free energy, and higher order
($|\psi|^6$ and beyond) terms need to be kept because very often the
quartic terms make {\em negative} contributions. Fortunately, the
complicated high LL wave functions have been studied in great detail
in the context of quantum Hall effect\cite{girvin}. In Ref.
\onlinecite{ym} we have used techniques developed in the studies of
quantum Hall effect to advance a very efficient method to evaluate
the free energy (\ref{F}) for the high LL wave function $\psi$, and
minimize it to determine the optimal VLS. The method is somewhat
technical and we refer the readers to Ref. \onlinecite{ym} for
details. More importantly, from the details of the VLS one can
extract the LL index $n$, from which we can get an {\em estimate} of
the order parameter momentum: $q\approx\sqrt{2n+1}/\ell$.

{\bf Spectra of Andreev Surface Bound State of d-wave FFLO
Superconductors Probed by Tunneling} --- The idea here is specific
to d-wave superconductors, which CeCoIn$_5$ is believed to be. In a
d-wave superconductor, the sign of the pairing order parameter
depends on the direction. As a consequence of this there exist
low-energy quasiparticle states that are bound to the surface of a
superconductor\cite{hu}. These so-called Andreev surface bound
states (ASBS) result from the change of sign of pairing order
parameter when a quasiparticle bounces off the surface; they give
rise to a zero bias conductance peak (ZBCP)\cite{hu,tanaka} in the
tunneling spectrum between a normal metal and the d-wave
superconductor (with proper orientation), separated by a potential
barrier. The ZBCP was recently observed in CeCoIn$_5$\cite{wei}. In
Ref. \onlinecite{cui}, we find the spectrum of ASBS changes when the
d-wave superconductor is driven into the FFLO state, and depends on
the momentum of the pairing order parameter. In particular, this
leads to a shift and split of the ZBCP in the tunneling spectrum,
with the split proportional to the order parameter momentum. This
provides yet another way to measure the order parameter momentum
using tunneling.

{\bf Other Possible Experiments} --- In Ref. \onlinecite{bbm},
Bulaevskii and coworkers proposed using interlayer transport in
quasi-2D superconductors in the presence of an in-plane magnetic
field to detect the FFLO state. The idea bears some similarity to
that of Ref. \onlinecite{kun2000}: in the superconducting phase,
interlayer transport is dominated by Josephson tunneling; for the
FFLO state the order parameter has spatial modulation, and the
Josephson effect is enhanced when the order parameter modulation is
commensurate with the phase modulation of the interlayer Josephson
coupling due to the in-plane field. The authors of Ref.
\onlinecite{bbm} have worked out the commensuration condition based
on certain assumption on the spatial structure of the order
parameter, under which the interlayer transport is enhanced (i.e.,
enhanced critical current or conductance). Experimentally one can
tune the in-plane magnetic field to look for such enhancement
associated with the commensuration, from which the wave vector of
the order parameter may be extracted. This experiment is also
``phase-sensitive".

In addition to probing the spatial structure of the superconducting
order parameter directly using the experiments discussed above, one
can also try to probe the spatial distribution of the unpaired
spin-up electrons in the FFLO state, which is closely related to the
order parameter structure. For example, for the one-dimensional,
Larkin-Ovchinnikov type order parameter structure (or the smectic
phase), one expect the unpaired spin-up electrons to localize along
the domain walls where the order parameter changes sign; thus the
periodicity (or wave length) of the spin modulation should be
one-half of that of the superconducting order parameter. The spatial
structure of the spins can be detected from elastic neutron or muon
scattering experiments. While such experiments do not directly probe
the order parameter structure and are thus not ``phase-sensitive",
being able to detecting the spin structure should also provide
convincing evidence for the FFLO state, and allow us to extract the
order parameter structure from it.

\subsection{Detection of Novel Pairing Phases in Cold Atom Systems}

As discussed in Section I, recent experiments\cite{zwierlein,randy}
have started to explore trapped cold atom systems with pairing
between atoms of different species (or hyperfine quantum number) and
unequal densities. Such systems have also generated strong
theoretical interest
recently\cite{combescot,mizushima,yangprl05,sedrakian,son,sheehy,theory,kinnunen}.
In particular, the possibility of realizing the FFLO state has been
discussed, and in Ref. \onlinecite{mizushima} it was suggested that
it can be detected by imaging the density profiles of each of the
pairing species, which should be oscillatory in real space for the
FFLO state. In another paper\cite{kinnunen}, it was suggested that
radio-frequency spectroscopy can be used to detect both phase
separation and the FFLO state. In Ref. \onlinecite{yangprl05} we
proposed two alternative methods to detect the FFLO state, which
directly probes the momenta of the Cooper pairs, using the methods
advanced in Refs. \onlinecite{regal,altman,greiner}. In Ref.
\onlinecite{regal} one projects the Cooper pairs of a BCS state onto
molecules by sweeping the tuning field through the Feshbach
resonance, and then removes the trap and uses time-of-flight (TOF)
measurement to determine the molecular velocity distribution and the
condensate fraction. One can do exactly the same experiment on the
FFLO state; the fundamental difference here is that in this case
because the Cooper pairs carry intrinsic (non-zero) momenta, the
condensate will show up as peaks corresponding to a set of {\em
finite} velocities in the distribution. Another method to detect the
Cooper pairs is to study the correlation in the shot noise of the
fermion absorption images in TOF\cite{greiner}, first proposed in
Ref. \onlinecite{altman}. In Ref. \onlinecite{greiner} the shot
noise correlation clearly demonstrates correlation in the occupation
of ${\bf k}$ and $-{\bf k}$ states in momentum space when weakly
bound diatom molecules are dissociated and the trap is removed. In
principle the same measurement can be performed on fermionic
superfluid states, and for an FFLO state, it would reveal
correlation in the occupation of ${\bf k}$ and $-{\bf k}+{\bf q}$
states, where ${\bf q}$ is one of the momenta of the pairing order
parameter\cite{yangprl05}. Both methods allow one to directly
measure ${\bf q}$, which defines the FFLO state. These methods are
unique to the cold atom systems; very similar ideas have also been
discussed in Ref. \onlinecite{sheehy}.

As discussed in section II, in addition to the spatial structure of
the pairing order parameter, we also need to detect the structure
and in particular the topology of the Fermi surface(s) of the
unpaired fermions. As discussed in Ref. \onlinecite{sy}, this can be
detected from the momentum distributions of the atoms, using TOF
after removing the trap. Such a measurement was recently performed
in a gas of $^{40}$K across a Feshbach resonance\cite{regal05}, and
hopefully will be performed in systems with unbalanced pairing in
the future. In the experiment of Ref. \onlinecite{regal05} the
effect of the trap on the momentum distribution appears to be quite
strong, such that the discontinuity in momentum distribution gets
wiped out even for non-interacting fermions. We hope that by
manipulating the form of the trap potential, its effect can be
minimized so that discontinuities in momentum distribution
associated with Fermi surfaces can be detected in future
experiments; this would probably require a trap potential that is
flat inside the trap and rises very fast near the boundary. It has
also been pointed out\cite{sedrakian} that in the deformed Fermi
surface pairing state, the TOF experiment will find anisotropy in
the distribution of the fermion velocity. Again one needs to
carefully analyze the effect of the trapping potential in this case.

We note in passing that the possibility of detecting some of the
novel unbalanced phases in nuclear matter has been discussed
recently\cite{isayev}, and the possibility that at large imbalance
the system may switch from s-wave to p-wave pairing\cite{bulgac},
including its detection.

\section{Summary}

In this chapter we have discussed how to characterize and detect
various possible phases that may result from pairing fermions with
different species and density imbalance. We argue in section II that
all the possible phases may be completely characterized by (i) the
spatial structure of the pairing order parameter; and (ii) the
structure and in particular, topology of the Fermi surfaces formed
by unpaired fermions.

These novel pairing phases may be realized in spin-singlet
superconductors subject to a Zeeman splitting between electron spin
states (either due to an external magnetic field or spontaneous
magnetization), trapped cold atom systems, and high density
quark/nuclear matter. For superconductors, the best case so far is a
quasi-two-dimensional heavy fermion superconductor CeCoIn$_5$, where
evidence for the realization of FFLO state has been found when it is
subject to a large in-plane magnetic field. While the existing
evidence from various experiments are quite strong, they are all
circumstantial in the sense that they do not directly probe the
spatial structure of the pairing order parameter. We hope some of
the ``phase sensitive" experiments we discussed in section IIIA will
lead to definitive proof of the FFLO state in this or other
superconductors. Experimental work on unbalanced pairing in trapped
cold atom systems have just started. Thus far clear evidence of
phase separation\cite{bedaque} between paired and unpaired fermions
have been found\cite{zwierlein,randy} when the imbalance is large.
Further work is needed to clarify whether some of the novel pairing
phases discussed in this and other chapters in this book are
realized at small imbalance, and the methods discussed in section
IIIB will hopefully be useful in that task.

\acknowledgments Over a period of nearly ten years, the author has
benefitted greatly from collaborations with Dan Agterberg, Qinghong
Cui, Denis Dalidovich, Chia-Ren Hu, Allan MacDonald, Subir Sachdev,
Shivaji Sondhi, and John Wei on the problem of pairing between
unbalanced fermion species. His work on this subject has been
supported by National Science Foundation grants DMR-9971541 and
DMR-0225698, as well as the Alfred P. Sloan Foundation and the
Research Corporation. This chapter was written while the author was
on sabbatical leave and visiting Harvard University and University
of California at Los Angeles; he thanks Professors Subir Sachdev and
Sudip Chakravarty for their warm hospitality, and acknowledges
partial support from a Florida State University Research Foundation
Cornerstone grant during his sabbatical leave.


\end{document}